\documentclass[12pt]{iopart}

\begin{document}

\title{Evaluation of the matrix exponential function using finite elements in time}

\author{D H Gebremedhin, C A Weatherford, X Zhang,  A Wynn III, G Tanaka}

\address{Department of Physics, Florida A \& M University, Tallahassee, FL 32307, USA}
\ead{charles.weatherford@famu.edu}
\begin{abstract}
The evaluation of a matrix exponential function is a classic problem of computational linear algebra. Many different methods have been employed for its numerical evaluation [Moler C and van Loan C 1978 {\it SIAM Review}{\bf \ 20} 4], none of which produce a definitive algorithm which is broadly applicable and sufficiently accurate, as well as being reasonably fast. Herein, we employ a method which evaulates a matrix exponential as the solution to a first-order initial value problem in a fictitious time variable. The new aspect of the present implementation of this method is to use finite elements in the fictitious time variable. [Weatherford C A, Red E, and Wynn A 2002  {\it Journal of Molecular Structure}{\bf  \ 592} 47] Then using an expansion in a properly chosen time basis, we are able to make accurate calculations of the exponential of any given matrix as the solution to a set of simultaneous equations.
\end{abstract}


\section*{Introduction}

The evaluation of the exponential of a square matrix $\e^{\bf A}$ is a classic problem of computational linear algebra.[1] A large number of methods have
been proposed and used for its evaluation. None of these methods have produced a generally applicable method which is sufficiently accurate as well as being reasonably fast.
Thus this problem might be considered unsolved. It is clearly an important and pervasive problem which arises in a wide variety of contexts.[2,3] A method which is capable of producing
a solution to essentially arbitrary precision would thus be of great importance. The present work uses a variant of a method described in Ref. 1,  namely the introduction of an
artificial  time-parameter
which produces an initial-value problem.  Instead of calling a \lq canned\rq \ solver, the present work uses a method introduced by several of the present authors [4] to solve quantum mechanical
initial-value problems. In this method, a finite element technique is used to propagate from an initial condition at $t=0$, which is the unit matrix, to the desired result at $t=1$. The time axis
is broken up into an arbitrary number of time elements and the solution is propagated from element to element, using a special basis in time introduced here for the first time. 

The next section presents the analysis of the problem and describes the solution algorithm. Then the algorithm is applied to evaluate the exponential of a number of test matrices. Finally, the
conclusions are presented.

\section*{Analysis and Solution Algorithm}

The problem at hand is the evaluation of the exponential of a square (generally complex) matrix $e^{\bf A}$. The present method introduces an
artificial time parameter so as to transform the evaluation into the solution of an initial-value problem. For a given  $n\times n$ square matrix $\mathbf A$, consider the following parametrized function definition:

\begin{equation} 
   \mathbf{\Psi}(t) \equiv \rme^{\mathbf{A}t}, \\
\end{equation}

\noindent
where $\bf \Psi$ is also an $n\times n$ square matrix and  the desired solution is $\mathbf{\Psi}(1)=\rme^{\mathbf{A}}$ which evolves
from the initial value given by $\mathbf{\Psi}(0)=\mathbf{1}$ ($\mathbf{1}$ is the diagonal unit matrix). This is a solution of the following linear ordinary differential equation, written for each matrix element,
\begin{equation}
\dot{\Psi}_{ij}(t)=\sum_{k=1}^{n}A_{ik}\Psi_{kj}(t),
\end{equation}

\noindent
where the over-dot stands for the time-derivative. The time axis extends over the interval $[0, 1]$. Now break the time axis
up into elements that extend between nodes $t_{\rm{i}}$ and $t_{\rm{i+1}}$, and define a local time $\tau$ that spans $[-1, 1]$.
The local time transformation is defined by the relation, 
\begin{equation}
\tau =qt -p,
\end{equation}

\noindent
where, $q =  2/(t_{\rm{i+1}} - t_{\rm{i}}) $ and $p= (t_{\rm{i+1}} + t_{\rm{i}})/(t_{\rm{i+1}} - t_{\rm{i}})$.
Thus, for an arbitrary time element $e$, Eq. $(2)$ can be written in terms of local time $\tau$ as

\begin{equation}
q\dot{\Psi}^{(e)}_{ij}(\tau)=\sum_{k=1}^{n}A^{(e)}_{ik}\Psi^{(e)}_{kj}(\tau).
\end{equation}

At this point, we will use the following \emph{ansatz} for $\bf \Psi^{(e)}$ to enforce continuity between two consecutive finite elements

\begin{equation}
\Psi^{(e)}_{ij}(\tau) = f^{(e)}_{ij}(\tau)+\Psi^{(e-1)}_{ij}(+1), \qquad f^{(e)}_{ij}(-1) = 0
\end{equation}

\noindent
and expand $f^{(e)}_{ij}(\tau)$ as

\begin{equation}
f^{(e)}_{ij}(\tau) = \sum_{\mu=0}^{m-1}  B_{\mu}^{ij(e)} s_\mu (\tau)
\end{equation}

\noindent
in a basis we define by

\begin{equation}
s_{\mu}(\tau) = \int_{-1}^\tau T_{\mu}(\tau) \, \rm{d}\tau
\end{equation}

\noindent
where $T_{\mu}(\tau)$ are Chebyshev Polynomials of the first kind.[5] Note that these basis functions enforce the
initial condition on the $f$'s given in Eq. (5) since $s_{\mu}(-1)=0$. The result for the decomposition of $f$ in $m$ basis functions is

\begin{eqnarray}
\Psi^{(e)}_{ij}(\tau) = \sum_{\mu=0}^{m-1} B_{\mu}^{ij(e)} s_\mu (\tau) +\Psi_{ij}^{(e-1)}(+1) \\
\dot{\Psi}^{(e)}_{ij} (\tau) = \sum_{\mu=0}^{m-1} B_{\mu}^{ij(e)} T_\mu (\tau).
\end{eqnarray}

\noindent
Now, insert (8) and (9) into Eq. (4), and multiply from the left by $w(\tau) s_{\mu'}(\tau)$ and integrate from $-1$ to $+1$
(note that  $w(\tau) = (1- \tau^2)^{-1/2}$ is the weighting function for Chebyshev polynomials). Rearranging terms we get,

\begin{equation}
\fl \eqalign {q\sum_{\mu} [\int_{-1}^{1} s_{\mu'}(\tau) \omega (\tau) T_{\mu}(\tau) \, \rm{d} \tau] B_{\mu}^{ij(e)} =
\sum_{k\mu} A^{(e)}_{ik} [\int_{-1}^{1} s_{\mu'}(\tau) \omega (\tau) s_{\mu}(\tau) \, \rm{d} \tau] B_{\mu}^{kj(e)} \cr
+ \sum_{k} A^{(e)}_{ik} [\int_{-1}^{1} s_{\mu'}(\tau) \omega (\tau) T_{0}(\tau) \, \rm{d} \tau] \Psi_{kj}^{(e-1)}(+1)}
\end{equation}

\noindent
where, $T_{0}(\tau) = 1$. Defining, the integrals in the above equation as

\begin{eqnarray}
C_{\mu' \mu} &\equiv \int_{-1}^{1} s_{\mu'}(\tau) \omega (\tau) T_{\mu}(\tau) \, \rm{d} \tau \\
D_{\mu' \mu} &\equiv \int_{-1}^{1} s_{\mu'}(\tau) \omega (\tau) s_{\mu}(\tau) \, \rm{d} \tau \\
g_{\mu'} &\equiv \int_{-1}^{1} s_{\mu'}(\tau) \omega (\tau) T_{0}(\tau) \, \rm{d} \tau
\end{eqnarray}.

\noindent
and substituting Eqs. (11-13) into Eq. (10) gives,

\begin{equation}
{q\sum_{\mu} C_{\mu' \mu} B_{\mu}^{ij(e)} =
\sum_{k\mu} A^{(e)}_{ik} D_{\mu' \mu} B_{\mu}^{kj(e)}
+ g_{\mu'} \sum_{k} A^{(e)}_{ik} \Psi_{kj}^{(e-1)}(+1)}
\end{equation}

\noindent
or, rearranging

\begin{equation}
\sum_{\mu k} (q C_{\mu' \mu} \delta_{ik} - A^{(e)}_{ik} D_{\mu' \mu} )B_{\mu}^{kj(e)}
= g_{\mu'} \sum_{k} A^{(e)}_{ik} \Psi_{kj}^{(e-1)}(+1)
\end{equation}

\noindent
where $\delta_{ik}$ is the usual Kronecker delta function. Then rewrite Eq. (15) as

\begin{equation}
\sum_{\mu k} \Omega^{(e)}_{(\mu' i)(\mu k)}B_{\mu}^{kj(e)}
= \Gamma_{\mu'}^{ij(e,e-1)}
\end{equation}

\noindent
where

\begin{eqnarray}
\Omega^{(e)}_{(\mu' i)(\mu k)} &\equiv (q C_{\mu' \mu} \delta_{ik} - A^{(e)}_{ik} D_{\mu' \mu} ) \\
\Gamma_{\mu'}^{ij(e,e-1)} &\equiv g_{\mu'} \sum_{k} A^{(e)}_{ik} \Phi_{kj}^{(e-1)}(+1).
\end{eqnarray}

Equation (16) is a set of simultaneous equations of size $(n \times m)$, which can be written in matrix form as, 

\begin{equation}
\mathbf{\Omega^{(e)} B}^{j (e)}= \mathbf{\Gamma}^{j(e,e-1)}\qquad j = 1, 2,..., n.
\end{equation}

\noindent
Here, $\mathbf{\Omega^{(e)}}$ is a (complex) matrix and for each $j$, $\mathbf{\Gamma^{j(e,e-1)}}$ and $\mathbf{B^{j(e)}}$ are vectors. 
Eq. (19) applies for each time element $e$. The solution is propagated from element to element, from $t=0$ to $t=1$. 
The above equation can be solved numerically in many ways, but we have chosen the method of LU decomposition.[6]
The present method is ideally suited to high-performance computers where the solver of choice would probably
be iterative. In the present case, we
apply LU decomposition to 
$\mathbf{\Omega^{(e)}}$ and back substituting all of the $\mathbf{\Gamma}^{j(e,e-1)}$'s, we will have all the elements for the matrix (which can also be viewed as three dimensional) $\mathbf{B^{j(e)}}$. This LU decomposition only needs to be done once since $\mathbf{\Omega^{(e)}}$ is independent of time. Thus, the propagation
just involves a matrix vector multiply.
Then, we employ Eq. $(8)$ to solve for $\mathbf{\Psi^{(e)}}(\tau = 1)$ for the element e, which, in turn, will be used as $\mathbf{\Psi^{(e + 1)}}(\tau = -1)$ for the next element e + 1. Starting off with a unit matrix for $\mathbf{\Psi^{(1)}}(t = 0)$, we continue this process till we calculate $ \mathbf{\Psi(t = 1)}$ at the last node, which is the exponential of the given matrix $\mathbf{A}$.

\section*{Results}

The calculations presented below were done on a Macintosh Intel laptop using Gnu C++ which has machine accuracy limit of $2.22045\times 10^{-16} $.
As an illustration, let's borrow a 'pathological' matrix from [1], which we have modified slightly to make it even worse. Consider a matrix $\mathbf{M1}$ given by,

\begin{eqnarray}
\eqalign{\mathbf{M1} &= \left[\begin{array}{cc}-73 & 36 \\-96 & 47\end{array}\right] \\
&=\left[\begin{array}{cc}1 & 3 \\2 & 4\end{array}\right] \left[\begin{array}{cc}-1 & 0 \\0 & -25\end{array}\right]  {\left[\begin{array}{cc}1 & 3 \\2 & 4\end{array}\right]}^{-1}.}
\end{eqnarray}

\noindent
The exponent of $\mathbf{M1}$ can be easily calculated as,
\begin{eqnarray*}
e^\mathbf{M1} &= \left[\begin{array}{cc}1 & 3 \\2 & 4\end{array}\right] \left[\begin{array}{cc}e^{-1} & 0 \\0 & e^{-25}\end{array}\right]  \left[\begin{array}{cc}-2 & 3/2 \\1 & -1/2\end{array}\right] \\
&= \left[\begin{array}{cc}{-2e^{-1}+3e^{-25}} & {(3/2)(e^{-1}-e^{-25})} \\{-4e^{-1}+4e^{-25}} & {3e^{-1}-2e^{-25}}\end{array}\right].
\end{eqnarray*}

\noindent
The above matrix, exact to $16$ decimal places, is given by

\begin{equation}
e^\mathbf{M1}\cong \left[\begin{array}{cc}-0.7357588823012208 & 0.5518191617363316 \\-1.4715177646302175 & 1.1036383234865511\end{array}\right].
\end{equation}

\noindent
The result of our program is displayed below and we run it by using just $8$ time steps and $8$ basis functions. The result is accurate to $13$ decimal places already.

\begin{equation}
e^\mathbf{M1}\cong \left[\begin{array}{cc}-0.7357588823012(181) & 0.5518191617363(358) \\-1.4715177646302(120) & 1.1036383234865(592)\end{array}\right].
\end{equation}

As an example of a non - diagonalizable matrix, consider the following matrix $\mathbf{M2}$, with complex eigenvalues

\begin{equation}
\mathbf{M2} = \left[\begin{array}{cc}0 & -1 \\1 & 0\end{array}\right]
\end{equation}

\noindent
It can be shown that,

\begin{equation}
e^\mathbf{M2} = \left[\begin{array}{cc}cos(1) & -sin(1) \\sin(1) & cos(1)\end{array}\right]
\end{equation}

\noindent
We are able to achieve $14$ decimal digit accuracy with $8$ time steps and $8$ basis functions.

\begin{equation}
e^\mathbf{M2} \cong \left[\begin{array}{cc} 0.54030230586814 & -0.84147098480790 \\ 0.84147098480790 & 0.54030230586814\end{array}\right]
\end{equation}

\noindent
 Table 1 shows the minimum number of basis functions, for a given number of time steps, which were required to achieve a precision of $\pm 1\times 10^{-14}$ on matrices whose exponential is known exactly. The matrices chosen are: the simplest possible matrix - a $2 \times 2$ real unit matrix, for which the result is the constant $e$ on the diagonals, and the matrices $\mathbf {M1}$ and $\mathbf {M2}$.

Let's check our program on matrices, which we picked randomly and for which we had no $\it apriori$ knowledge as to the result of their exponentiation. We fixed 8 time steps and/or 8 basis functions, and varied the other corresponding parameter from 5 to 40 and checked how the results of the program varied in accuracy. For the sake of saving space, we only displayed the result of the last element--the other elements of the matrix exponential behaved similarly. 
The matrices chosen are a $5 \times 5$ real matrix $\mathbf{M3}$,

\begin{equation}
\mathbf{M3} = \left[\begin{array}{ccccc}-0.1 & -0.2 & -0.3 & -0.4 & -0.5 \\-0.6 & -0.7 & -0.8 & -0.9 & -1 \\0.1 & 0.2 & 0.3 & 0.4 & 0.5 \\0.6 & 0.7 & 0.8 & 0.9 & 1 \\1 & 2 & 3 & 4 & 0\end{array}\right]
\end{equation}

\noindent
and a $3 \times 3$ complex matrix $\mathbf{M4}$

\begin{equation}
\mathbf{M4} = \left[\begin{array}{ccc}1+i & 1-i & i \\1 & 2i & 0 \\1+2i & -1+i & -1-i\end{array}\right].
\end{equation}

From Table 2, one can see that the numbers up to $12$ decimal digits have saturated after $5$  time steps and/or basis functions. Similarly, Table 3 shows $13$ digits of accuracy as we switch the two parameters from $5$ to $40$, except for the case of $5$ basis functions, which only shows $8$ accurate significant digits. This shows that for complex matrices, there is inherently more work for the program to handle because of the imaginary part of the matrix elements and there is apparently more sensitivity to the number of basis functions used than to the number of time steps.

\Table{\label{tone}Minimum number of basis functions and time steps required for a precision of  $\pm 1\times 10^{-14}$ for a $2 \times 2$ unit matrix, $\mathbf{M1}$ and $\mathbf{M2}$.}
\br
\centre{2}{$2\times 2$ unit matrix}&\centre{2}{$\mathbf{M1}$}&\centre{2}{$\mathbf{M2}$}\\

\crule{2}&\crule{2}&\crule{2}\\
Time steps & Basis functions& Time steps & Basis functions& Time steps & Basis functions\\
\mr
\01&	11&	\0\005&		-&		\01&		11\\
\02&	\09&	\0\008&		7&	\02&		\09\\
\04&	\08&	\016&		6&	\04&		\08\\
\08&	\07&	\050&		5&	\08&		\07\\
16&	\06&	256&			4&	15&		\06\\
58&	\05&	\0\0-&		-&		40&		\05\\
\br
\end{tabular}
\end{indented}
\end{table}

\Table{\label{ttwo}Results of matrix $e^{M3}_{5 5}$ for typical runs of 8 time steps and 8 basis functions.}
\br
Result& Time steps & Basis functions\\
\mr
\underline{3.210309305973}118 &		\05&		\08\\
\underline{3.210309305973}288 &		40&		\08\\
\underline{3.210309315373}377&		\08&		\05\\
\underline{3.210309305973}281&		\08&		40\\
\br
\end{tabular}
\end{indented}
\end{table}

\Table{\label{tthree}Results of matrix $e^{M4}_{3 3}$ for typical runs of 8 time steps and 8 basis functions.}
\br
Result& Time steps & Basis functions\\
\mr
\underline{-0.5119771222980}63 - i \underline{0.0897728113135}12 &		\05&		\08\\
\underline{-0.5119771222980}81 - i \underline{0.0897728113135}26 &		40&		\08\\
\underline{-0.51197712}1264660 - i \underline{0.08977281}0979965&		\08&		\05\\
\underline{-0.5119771222980}82 - i \underline{0.0897728113135}26&		\08&		40\\
\br
\end{tabular}
\end{indented}
\end{table}

\section*{Conclusion}

We have presented a robust, easily used, and accurate algorithm for the evaluation of the exponential of a matrix. We did this by introducing an
artificial time parameter and evaluating the matrix exponential as the solution of an initial-value problem in this artificial time. We solved the initial-value problem
by using finite elements in time with a new time basis which we defined here so as to enforce the initial conditions on the solution at the beginning of each time finite element. This resulted in set of simultaneous equations for the expansion coefficients. 
The actual algorithm employed here was an LU decomposition which was very fast and efficient. The relative efficiency of the method should be most
apparent when implemented on high-performance computers since the algorithm is highly parallel.
The method was applied to several matrices as a proof of the validity of the algorithm. 
The results of our calculations show that we only need about $8$ basis functions and $8$ time steps for the matrices considered for accuracies as great as $13$ significant digits. We trust that this method of numerically  calculating the exponential of a matrix will be recognized  to be a \emph{nondubious} one! \\

\section*{Acknowlegements}
This work was supported by the NSF CREST Center for Astrophysical Science and Technology under Cooperative Agreement 
HRD-0630370.

\end{document}